\documentclass[10pt]{iopart}
\usepackage{iopams} 
\usepackage{graphicx}

\def\apj{{\itshape Astrophys. J.}}
\def\aap{{\itshape A\&A }}
\def\prl{{\itshape Phys. Rev. Lett.}}
\def\mnras{{\itshape Mon. Not. R. Astron. Soc.}}
\def\ang{{\itshape Ann. Geophys.}}

\def\jgr{{\itshape J. Geophys. Res.}}
\def\pop{{\itshape Phys. Plasmas}}
\def\ppcf{{\itshape Plasma Phys. Control. Fusion}}
\def\ssr{{\itshape Space. Sci. Rev. }}
\def\mnras{{\itshape Mon. Not. R. Astron. Soc.}}

\begin{document}

\title{Dissipation in Relativistic Pair-plasma Reconnection: Revisited}

\author{Seiji Zenitani}

\address{Research Institute for Sustainable Humanosphere, Kyoto University, Gokasho, Uji, Kyoto 611-0011, Japan}
\ead{zenitani@rish.kyoto-u.ac.jp}

\begin{abstract}
Basic properties of relativistic magnetic reconnection
in electron--positron pair plasmas
are investigated by using a particle-in-cell (PIC) simulation.
We first revisit a problem by Hesse \& Zenitani (2007),
who examined the kinetic Ohm's law across the X line.
We formulate a relativistic Ohm's law
by decomposing the stress-energy tensor. 
Then, the role of the new term, called the heat-flow inertial term,
is examined in the PIC simulation data.
We further evaluate the energy balance in the reconnection system.
These analyses demonstrate physically transparent ways
to diagnose relativistic kinetic data.
\end{abstract}

%
%
%
%

\section{Introduction}

Magnetic reconnection is a fundamental plasma process
to drive explosive events in space and astrophysical environments \cite{vas75}.
To allow the change in the field-line topology,
the ideal condition for arbitrary plasma species $s$,
\begin{equation}
\label{eq:E}
\vec{E} + \vec{V}_s \times \vec{B} = 0,
\end{equation}
should be violated around the X line,
across which the magnetic flux is transported.
In a collisionless plasma,
the central problems in reconnection physics are
how the ideal condition is violated and
what kinetic mechanisms are responsible for the violation.
This issue is often referred to as the dissipation problem,
because the violation of the ideal condition is necessary for
the nonideal energy dissipation that drives magnetic reconnection.

In a nonrelativistic plasma,
significant efforts have been paid to the electron idealness,
because the electrons are the last species to
decouple from the field lines near the X line.
The electron Ohm's law, obtained from the electron momentum equation, is
useful to analyze the problem,
\begin{equation}
\label{eq:nonrela}
\vec{E} + \vec{V}_e \times \vec{B} = 
- \frac{1}{e n_e} \nabla\cdot \mathbb{P}_e
- \frac{m_e}{e} \Big( \frac{\partial\vec{V}_e}{\partial t} + (\vec{V}_e \cdot \nabla) \vec{V}_e \Big),
\end{equation}
where $\mathbb{P}_e$ is the electron pressure tensor.
In two-dimensional quasisteady systems,
only the first terms in the left and right hand sides survive at the X line,
and therefore
the reconnection electric field is balanced by
the divergence of the electron pressure tensor term \cite{vas75}.
Cai \& Lee \cite{cai97} showed that
the pressure-tensor term balances the reconnection electric field
on a sub-ion timescale in PIC simulations.
Hesse et~al. \cite{hesse99} have demonstrated
with PIC simulations that
the divergence of the off-diagonal parts of the electron pressure tensor sustains
the reconnection electric field in a quasisteady stage.
He and his colleagues have long studied electron kinetic physics
that is responsible for the off-diagonal parts of the electron pressure tensor.
Insights from their works were reviewed by Ref.~\cite{hesse11}.

Magnetic reconnection is also believed to occur
in a relativistic plasma in high-energy astrophysical settings \cite{uz11a}.
The question then arises:
how the reconnection proceeds in a relativistic kinetic plasma?
However, since relativistic plasma physics is far more complicated than nonrelativistic one,
even an appropriate form of Ohm's law was not clear.
Theorists have constructed several relativistic forms of
Ohm's law \cite{bf93,gedalin96,meier04}
from fluid equations or multi-species Boltzmann equations.

Over the last decades,
kinetic modeling of relativistic magnetic reconnection have been in progress
\cite{zeni01,claus04,zeni07,zeni08,hesse07,liu11,bessho12,cerutti13,sironi14,melzani14,liuy15,werner16b}.
Although many of these studies focused on
the stability of a current-sheet, particle acceleration, and radiative signatures,
Hesse \& Zenitani \cite{hesse07} made the first attempt to
examine the kinetic Ohm's law near the X line in PIC simulations.
Employing Wright \& Hadley \cite{wright75}'s definition of the pressure tensor,
they formulated a relativistic Ohm's law from the Vlasov equation.
Then, they analyzed PIC simulation data of
magnetic reconnection in a relativistic pair plasma, and
found that the divergence of the ``pressure tensor'' term
sustains the reconnection electric field at the X line,
as in nonrelativistic cases.
This or equivalent analysis has been employed by
subsequent studies \cite{liu11,bessho12,melzani14,liuy15}.
However, there were some problems in the framework in Ref.~\cite{hesse07}.
First, the Wright--Hadley pressure tensor \cite{wright75} was not symmetric.
An asymmetric pressure tensor is not the one that we were looking for.
Second, physical meanings of the terms in Ohm's law were not adequately given.

Another important issue in reconnection physics
is the energy balance around the reconnection site.
Although this is a very fundamental problem,
scientists have started to discuss the energy balance very recently.
Using MHD simulations,
Birn et~al. \cite{birn05,birn09} have argued that
the reconnection process converts
the incoming magnetic energy flux into
the enthalpy flux in the outflow region.
Successive hybrid and PIC simulations \cite{aunai11c,yamada15,yamada16,lapenta17}
reported that the energy is mostly transferred to the enthalpy flux
in the kinetic regimes as well.
In a relativistic plasma, 
using relativistic two-fluid simulations,
Zenitani et~al.~\cite{zeni09b} demonstrated that
the enthalpy flux carries most of the outgoing energy
in antiparallel reconnection
and that
Poynting flux replaces the enthalpy flux
in the presence of the out-of-plane guide field.
To the best of our knowledge,
no one has studied the energy balance in relativistic reconnection
in depth with first-principle PIC simulations,
because researchers focused on other important issues
and
because the analysis framework was not well known. 

The purpose of this paper is
to re-examine the Ohm's law problem in Ref.~\cite{hesse07},
based on our up-to-date knowledge.
Decomposing the stress-energy tensor,
we derive a relativistic kinetic form of Ohm's law,
which contains a ``heat-flow inertial term.''
We further apply the decomposition to
the energy balance problem around the reconnection site.

This paper is organized as follows.
In Section \ref{sec:theory},
we describe basic issues of relativistic mechanics and then
we derive a relativistic extension of the electron Ohm's law.
In Section \ref{sec:numerical}, we describe the setup of our PIC simulation.
In Section \ref{sec:results}, we present the simulation results.
The relativistic Ohm's law is evaluated.
In Section \ref{sec:balance}, we further extend our discussion to
the energy balance problem during magnetic reconnection.
Section \ref{sec:discussion} contains discussions and a summary.


\section{Preparation}
\label{sec:theory}

Let us briefly review basic issues of relativistic statistical mechanics.
The metric tensor is set to $g^{\alpha\beta}={\rm diag}(-1,1,1,1)$.
We set $c=1$ throughout the paper, but
we sometimes keep $c$ to emphasize its physical meaning.
The particle flux density $N^{\alpha}$ and
the stress-energy tensor $T^{\alpha\beta}$ are defined in the following way,
\begin{equation}
N^{\alpha} = \int f(u) u^{\alpha} \frac{d^3u}{\gamma},
\label{eq:N}
~~~
T^{\alpha\beta} = \int f(u) u^{\alpha} u^{\beta} \frac{d^3u}{\gamma}
.
\label{eq:T}
\end{equation}
We express the properties in the other frame by the prime sign ($'$). 
Since ${d^3u}/{\gamma} = {d^3u'}/{\gamma'}$,
both follow the Lorentz transformation,
\begin{equation}
\label{eq:LT}
N^{\alpha} = \Lambda^{\alpha}_{\mu} N'^{\mu},
\label{eq:N2}
~~~
T^{\alpha\beta} = \Lambda_{\mu}^{\alpha} \Lambda_{\nu}^{\beta} T'^{\mu\nu}
\label{eq:T2}
\end{equation}
where $\Lambda$ is the Lorentz transformation tensor.

The second-rank tensor $T$ can be decomposed in the following way \cite{ec40}:
\begin{equation}
\label{eq:eckart}
T^{\alpha\beta}=\mathcal{E} u^{\alpha}u^{\beta} + q^{\alpha}u^{\beta} + q^{\beta}u^{\alpha} + P^{\alpha\beta}
\end{equation}
where
$\mathcal{E} \equiv T^{\alpha\beta}u_{\alpha}u_{\beta}$,
$q^{\alpha} \equiv -\Delta^{\alpha}_{\beta} T^{\beta\gamma} u_{\gamma}$,
$P^{\alpha\beta} \equiv \Delta^{\alpha}_{\gamma}\Delta^{\beta}_{\delta}T^{\gamma\delta}$, and
$\Delta^{\alpha\beta} = g^{\alpha\beta} + u^{\alpha}u^{\beta}$ is the projection operator.
Here, $u^{\alpha}$ is an arbitrary flow, 
$\mathcal{E}$ is the energy density in the $u^{\alpha}$-moving frame,
$q^{\alpha}$ and $P^{\alpha\beta}$ are projections of
the energy flux and the pressure tensor in the flow frame.
The four-vector $q^\alpha$ is called the heat flow.
We expect $P^{\alpha\beta}=pu^{\alpha}u^{\beta}+pg^{\alpha\beta}$ for an ideal gas,
where $p$ is a scalar pressure.

\begin{figure}[htbp]
\centering
\includegraphics[width={0.85\columnwidth},clip]{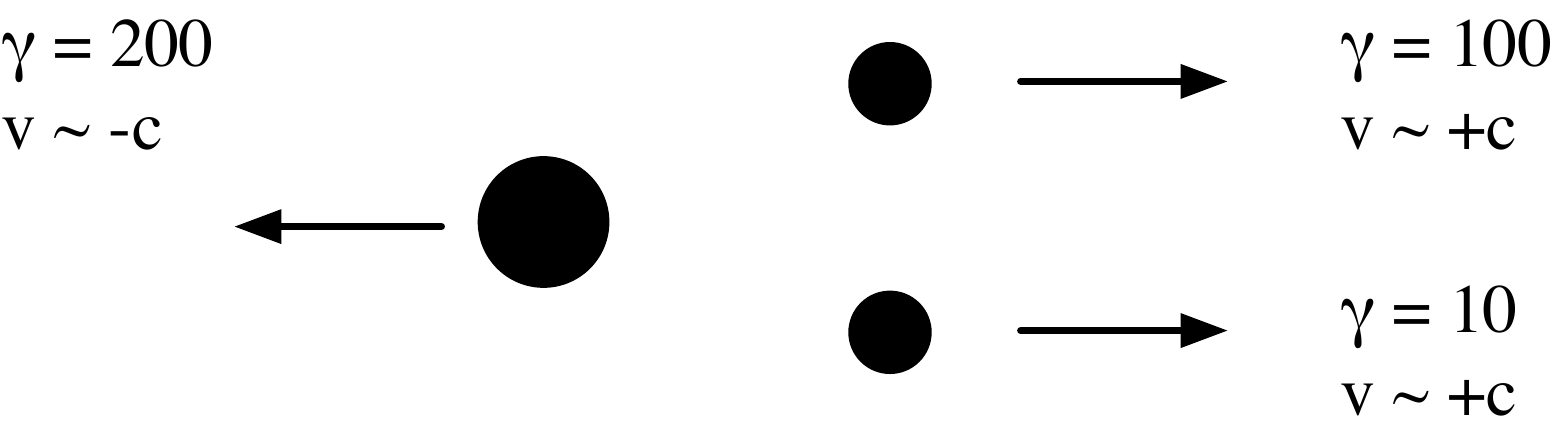}
\caption{
An example of a relativistic fluid system,
which consists of three particles.
\label{fig:cartoon}}
\end{figure}

This decomposition works for
an arbitrary four-velocity $u^{\alpha}$.
For practical use,
we need to choose an appropriate rest frame for the plasma flow.
However, unlike in the nonrelativistic case,
it is not straightforward to define the rest frame in a relativistic plasma.
This is schematically illustrated in Figure \ref{fig:cartoon}.
We assume that three particles travel in the $\pm x$ directions at relativistic speeds.
Their speeds are $\sim c$, but they have different Lorentz factors. 
In this case, an average plasma number flow should be in the right direction,
while the energy obviously flows in the left direction.
This tells us that there could be two rest frames,
in which the plasma number flux is zero, 
and
in which the plasma energy flux is zero.
The former is called the Eckart frame \cite{ec40},
and
the latter is called the Landau frame \cite{landau}.

The Eckart-frame four-velocity $u^\mu_{(E)}$ can be straightforwardly obtained
from the plasma number flux,
\begin{equation}
u^{\mu}_{(E)} = N^\mu / \sqrt{-N^{\nu}N_{\nu}} = N^\mu / n
,
\end{equation}
where $n=\sqrt{-N^{\nu}N_\nu}$ is the proper density.
The Landau-frame four-velocity $u^\mu_{(L)}$ can be obtained
from an eigenvalue problem of the stress-energy tensor \cite{synge56,werner16b}.
Our derivation will be presented in \ref{sec:Landau}.
The Eckart and Landau frames are identical in the nonrelativistic regime,
but it is important to distinguish the two frames in a relativistic plasma.
In this paper, we employ the Eckart frame as the fluid rest frame.
We assume $u^\mu=u^\mu_{(E)}$ unless stated otherwise.



We consider the energy momentum equation of the electron fluid,
\begin{equation}
\label{eq:T}
\partial_{\beta} T_{(e)}^{\alpha\beta}
= -e F^{\alpha\beta} N_{\beta}
= -e n F^{\alpha\beta} u_{\beta}
\end{equation}
where $F^{\alpha\beta}$ is the electromagnetic tensor.
From the momentum ($i\beta$) part of the equation,
we obtain a relativistic generalization of the electron Ohm's law.
\begin{eqnarray}
\label{eq:ohm}
\vec{E}
&=&
- \vec{V} \times \vec{B}
- \frac{1}{\gamma e n}
\Big(
\partial_t T^{i0}
+
\nabla \cdot ( \mathcal{E} u^iu^j + Q^{ij} + P^{ij} )
\Big)
 \\
\label{eq:ohm2}
&=&
- \vec{V} \times \vec{B}
- \frac{1}{\gamma e n}
\Big(
\partial_t ( \gamma \mathcal{E} u^i + Q^{i0} + P^{i0} )
\nonumber \\
&&
~~~~~~~~~~~~~~~~~~~~~~~
+
\nabla \cdot ( \mathcal{E} u^i u^j + Q^{ij} + P^{ij} )
\Big)
.
\end{eqnarray}
Here, $\vec{V}$ is the 3-velocity
for the electron Eckart velocity
and
${Q}^{\alpha\beta} \equiv q^{\alpha}u^{\beta} + q^{\beta}u^{\alpha}$ is 
the heat-flow part of the stress-energy tensor (Eq.~\ref{eq:eckart}).
In the detailed form (Eq.~\ref{eq:ohm2}),
the two $\mathcal{E}$-related terms of the right hand side
correspond to the bulk inertial effect.
They contain the effect of the relativistic gas temperature,
because the energy is equivalent to the mass.
For example, the term can be further split into
the conventional part and an additional part by the relativistically hot gas, i.e.,
$\mathcal{E} u^\alpha u^\beta = n u^\alpha u^\beta + (\mathcal{E}-n) u^\alpha u^\beta$.
The relativistic effects are included in
(1) the temperature part of the bulk inertial term,
(2) the heat-flow terms,
(3) the time derivative of these terms, and
(4) the Lorentz factors.

%
%
%

\section{Numerical model}
\label{sec:numerical}

We carry out a two-dimensional particle-in-cell (PIC) simulation of
relativistic magnetic reconnection.
We consider a pair plasma of electrons and positrons.
The Harris--Hoh current sheet is employed as an initial model \cite{harris,hoh66},
\begin{equation}
{\boldmath{B}} = B_0 \tanh([z-z_0]/L) {\boldmath\hat{x}}, 
\end{equation}
\begin{eqnarray}
f_{s}(\vec{u}) = \frac{n_0\cosh^{-2}([z-z_0]/L)}{4\pi m^2cT K_2(mc^2/T)}
\exp\big[
\frac{ -\gamma_s (\varepsilon - \beta_s mc u_y) }{ T }
\big] \nonumber \\
+ \frac{n_{bg}}{4\pi m^2cT K_2(mc^2/T)} \exp\big[-\frac{\varepsilon}{T} \big],
\end{eqnarray}
where the subscript $s$ denotes the species ($p$ for positrons and $e$ for electrons), the subscripts $0$ and $bg$ denote physical quantities
for the Harris-sheet plasmas and the uniform background plasmas,
$L$ is the thickness of the current sheet,
$K_2$ is the modified Bessel function of the second kind,
$T$ is the proper temperature,
$n_0$ is the proper density,
$\varepsilon=\gamma mc^2$ is the particle energy, and
$\beta_s$ is the drift speed of the Harris-sheet plasmas.
We set $m=1$, $\beta_{p,e}=\pm 0.3$, and $T=mc^2$.
The background density is set to $n_{bg} = 0.2 (\gamma_e n_0)$.
Using the cell size $\Delta$,
the Debye length and the current sheet thickness is set to
$3\Delta$ and $L=10\Delta$.
The time is normalized by the light crossing time of $L/c$.
The plasma beta in the inflow region is set to $0.2$.

The system size is $1000\Delta$ in $x$ and $ 600 \Delta$ in $z$.
We consider two current sheets in the periodic system, as in our previous study \cite{zeni07}.
The first Harris sheet is set at $z_0 = 450$
in the top half ($300 < z < 600$).
Then, the second Harris sheet is set at $z_0 = 150$
in the bottom half ($0 < z <300$).
The magnetic polarities and the electric current are oppositely set
for the second Harris sheet.
In the paper, we present the results in the top half.
Particles are initialized by a modified Sobol algorithm \cite{zeni15}.
Magnetic reconnection is triggered by
a small perturbation in a magnetic flux function.
The maximum amplitude of the perturbed magnetic field is $\lesssim 0.1 B_0$.

To better evaluate the stress-energy tensor $T^{\alpha\beta}$, 
we employ several workarounds.
First, we use a second-order shape factor in our PIC simulation.
Second, we use $2 \times 10^4$ pairs of particles per cell for the Harris-sheet density $\gamma_e n_0$.
Third, several properties are box-car averaged with neighboring cells.

\begin{figure}[htbp]
\centering
\includegraphics[width={\columnwidth},clip]{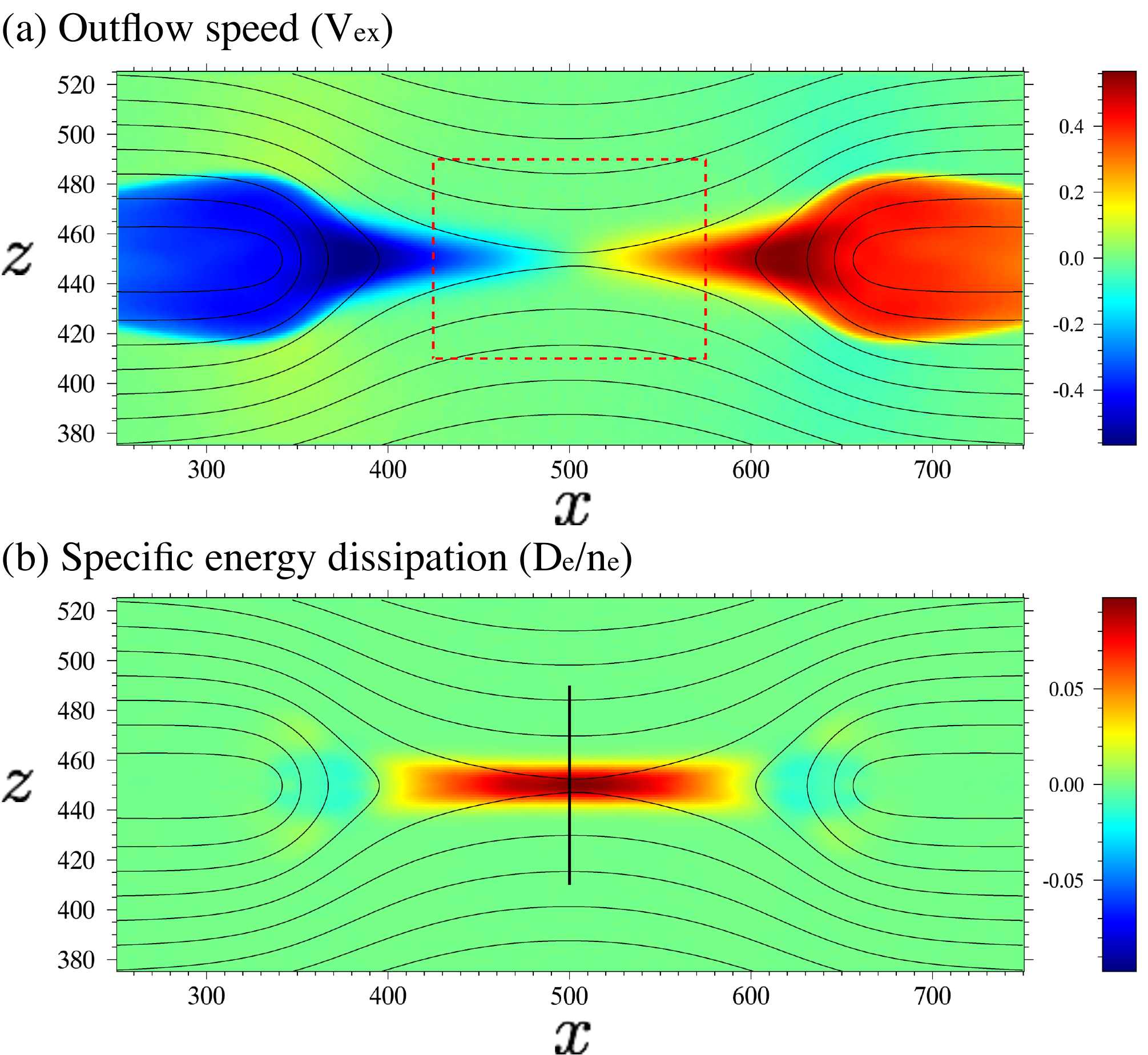}
\caption{
PIC simulation results at $t=85$.
(a) Electron outflow $V_{ex}$.
(b) Specific energy dissipation, $\mathcal{D}_e/n_e$ (Eq.~\ref{eq:DeN}).
\label{fig:overview}}
\end{figure}

\section{Results}
\label{sec:results}

The panels in Figure \ref{fig:overview} present
key properties at $t = 85$. 
Figure \ref{fig:overview}(a) shows
the outflow component of the electron flow, $V_{ex}$.
The bidirectional jets originate from the X line at the center.
The electron maximum speed is $\pm 0.57c$.
There are remnants of the initial current sheet,
at $x \lesssim 350$ and $x \gtrsim 650$.
The bottom panel shows
a specific energy dissipation ($\mathcal{D}_e/ n_e$),
\begin{equation}
\label{eq:DeN}
\frac{1}{ecB_0} \Big(\frac{\mathcal{D}_e}{n_e}\Big)
\equiv
\frac{
\gamma_e \big[ \vec{J} \cdot ( \vec{E} + \vec{V}_e \times \vec{B} ) - \rho_c \vec{V}_e \cdot \vec{E} \big]
}{en_e cB_0}
,
\end{equation}
where $\rho_c$ is the electric charge-density.
The $\mathcal{D}_e$ measure is proven to mark dissipation sites \cite{zeni11c}.
In this study, we further divide it by the electron proper density $n_e$,
in order to emphasize the dissipation region around the X line,
while keeping $\mathcal{D}_e/ n_e$ invariant.
Thanks to the normalization constant $(ec B_0)^{-1}$,
Equation~\ref{eq:DeN} conveniently approximates the reconnection rate,
$\mathcal{D}_e/ (en_ec B_0) \sim J_y E_y / (en_e c B_0) \sim ( E_y / B_0)$.
The maximum value $\approx 0.1$ around the X line is
comparable with the reconnection rate of this run.


\begin{figure}[htbp]
\centering
\includegraphics[width={\columnwidth},clip]{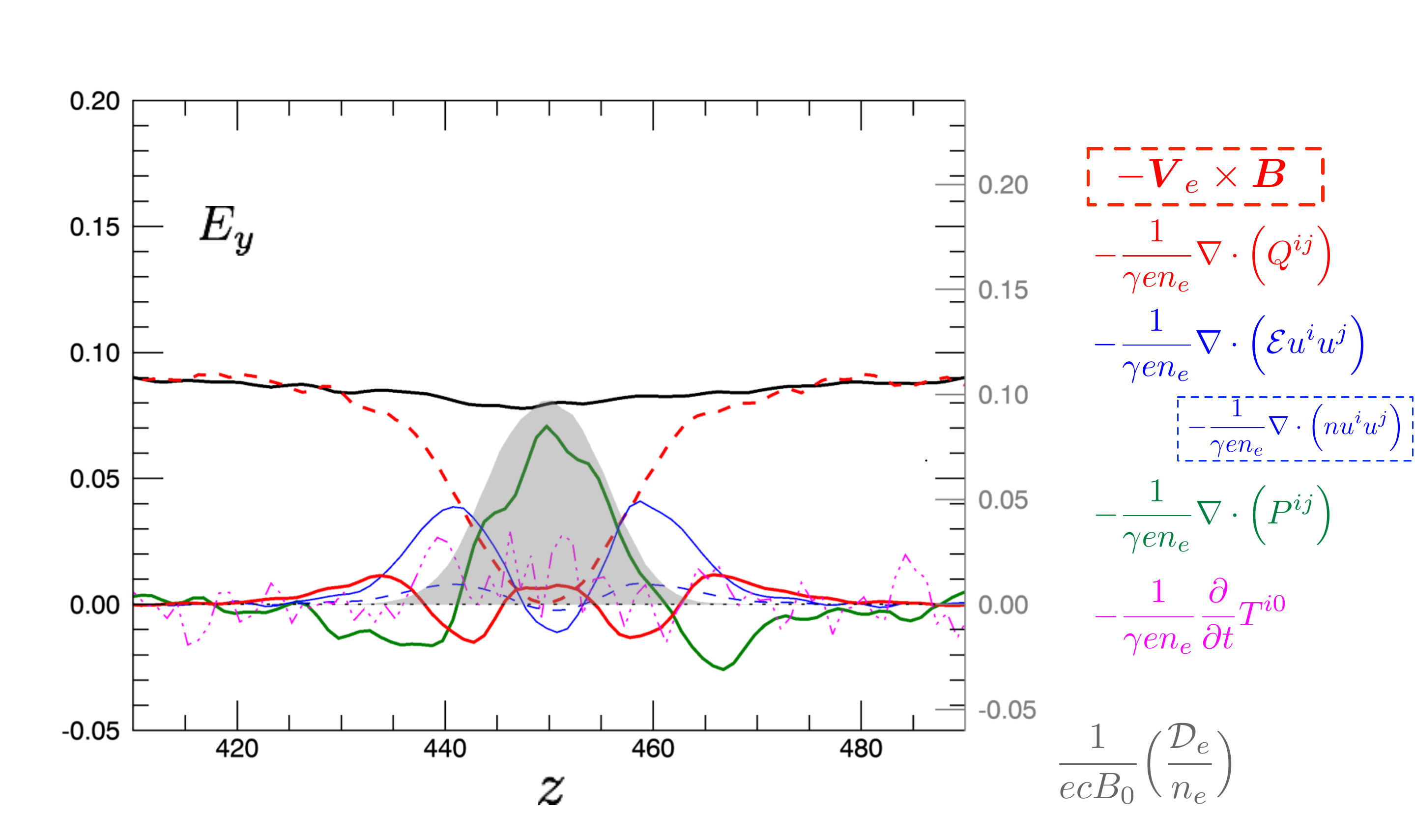}
\caption{
Composition of the reconnection electric field $E_y$ along the inflow line.
The terms are based on Equation \ref{eq:ohm}.
The gray shadow shows
the specific energy dissipation $\mathcal{D}_e/n_e$ (Eq.~\ref{eq:DeN}),
scaled by the right axis.
\label{fig:ohm}}
\end{figure}

Using the PIC simulation data,
we examine the violation of the electron idealness along the $x=500$ line,
as indicated by the black line in Figure \ref{fig:overview}(b).
Figure \ref{fig:ohm} shows
the composition of the reconnection electric field $E_y$,
based on Equation \ref{eq:ohm}.
The time derivative parts were combined in $\partial_t T^{i0}$,
because we are more interested in the divergence part,
which will persist in the quasisteady stage.
In Figure \ref{fig:ohm},
the reconnection electric field (the black line) is in good agreement with
the convection electric field (the red dashed line) in the inflow regions.
As we approach the center, the bulk inertial term (blue) increases,
which is partially canceled by the divergence of the pressure tensor (green).
The blue dashed line shows the proper-mass part of the bulk inertial term,
$\nabla \cdot (n_e u^iu^j)$.
This is much smaller than the entire bulk inertial term.
In this case,
the plasma temperature increases from $T \approx 1$ in the upstream region
to $\frac{1}{3}{\rm Tr}(P^{ij})/n \gtrsim 1.5$ near the center,
because plasmas are strongly energized around the reconnection site.
Consequently, the relativistic pressure substantially enhances
the bulk inertial effect. 
Near the center,
the bulk inertial term becomes unimportant and then
the pressure tensor term (green) is responsible for
the reconnection electric field $E_y$.
It is positive $-\frac{1}{\gamma e n_e} \nabla \cdot P^{ij}>0$ around $440 \lesssim z \lesssim 460$.
This corresponds to the local momentum transport
from the midplane $z \approx 450$. 
The central region corresponds to
the site of the enhanced energy dissipation.
The specific energy dissipation (Eq.~\ref{eq:DeN}) is indicated by
the gray shadow in Figure \ref{fig:ohm}.
This suggests that
the pressure-tensor term is the major contributor to
the energy dissipation $\mathcal{D}_e \approx \gamma_e J_y (\vec{E}+\vec{V}_e\times\vec{B})_y \approx -(J_y /{e n_e}) [\nabla \cdot (P^{ij})]_y$ near the X line. 
In addition, we find some contributions from
the temporal part of the inertial terms (magenta) and
the divergence of the heat-flow tensor term (red). 
In particular,
the heat-flow term only appears in the relativistic plasma.
Hereafter we call it the heat-flow inertial term.
Even though it is smaller than the pressure-tensor term,
we noticed that it persists throughout the simulation run.
It stems from the divergence of the off-diagonal parts of the electron heat-flow tensor,
in particular from $-\partial_z Q_{eyz}$.
Later in this section,
we will further discuss
the origin of the heat-flow inertial effect. 
 
\begin{figure}[htbp]
\centering
\includegraphics[width={\columnwidth},clip]{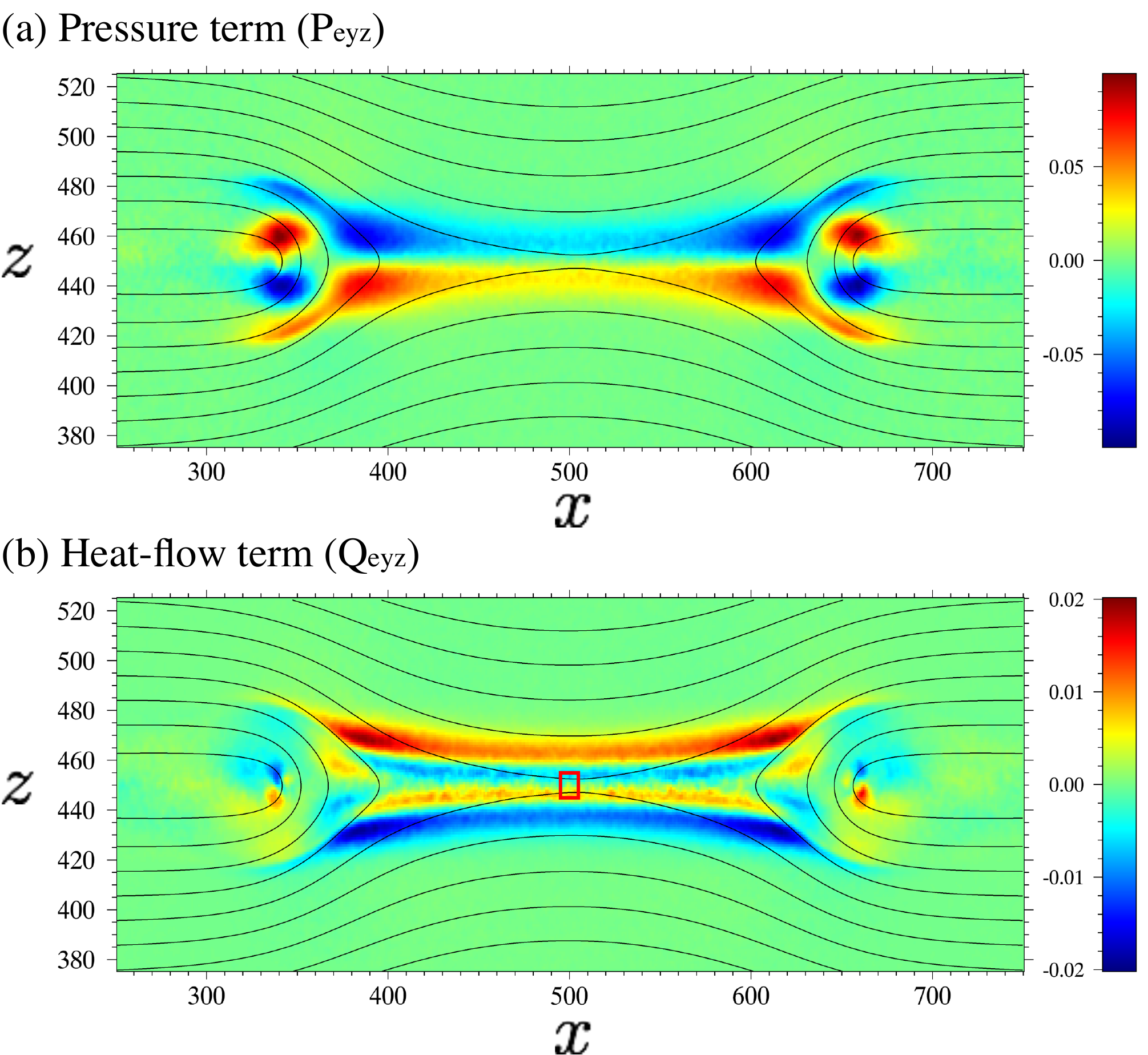}
\caption{
Spatial profiles of
(a) the pressure tensor term $P_{eyz}$ and
(b) the heat-flow term $Q_{eyz}$
at $t=85$.
\label{fig:PQ}}
\end{figure}

Shown in Figure \ref{fig:PQ} are
spatial profiles of $P_{eyz}$ and $Q_{eyz}$.
The local momentum transport $P_{eyz}$ has a bipolar structure
across the midplane (Fig.~\ref{fig:PQ}a).
In the reconnection layer, the electrons travel in $-y$.
Thus, the bipolar picture corresponds to
the diffusion of the electron current-carrying momentum
in the $\pm z$ directions, away from the midplane \cite{cai94,keizo09}.
This leads to the effective inertial effect via the $-\partial_z P_{eyz}$ term.
The heat-flow term $Q_{eyz}$ gives a more complicated picture (Fig.~\ref{fig:PQ}b).
It exhibits a 4-layer structure of
the outer bipolar layers and the inner bipolar layers. 
Around the midplane,
the inner layers have the same signs as
the bipolar $P_{eyz}$ layers.
The outer layers have the opposite signs from the inner layers.

\begin{figure}[htbp]
\centering
\includegraphics[width={\columnwidth},clip]{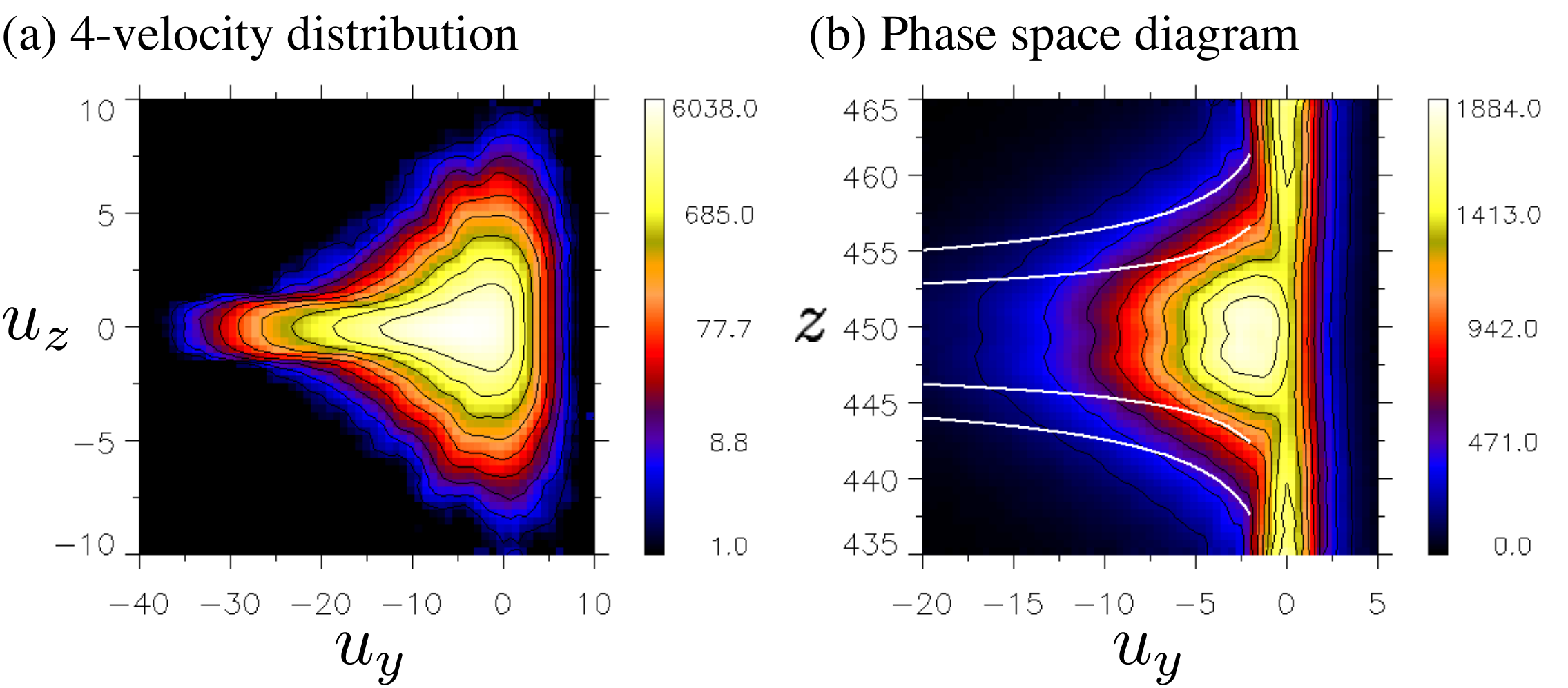}
\caption{
(a) Electron four-velocity distribution function ($u_y$--$u_z$) and
(b) electron phase-space diagram ($u_y$--$z$) at $t=85$.
The white lines indicate $|z-z_0| \propto \pm (-u_y)^{-1/3}$.
\label{fig:PSD}}
\end{figure}

To understand the problem from microscopic viewpoints,
we present an electron velocity distribution function (VDF) in $u_y$--$u_z$
around the X line in Figure \ref{fig:PSD}(a).
The electrons are sampled in the red box in Figure \ref{fig:PQ}(b)
($(x,z) \in [495,505] \times [445,455]$) at $t=85$.
The VDF is integrated in $u_x$. 
Although the VDF has a core component in the mildly relativistic domain
($-10 \lesssim u_y \lesssim 5$),
there exist highly-energetic electrons in the $-u_y$ direction,
down to $u_y \approx -35$. 
This is due to the direct $y$-acceleration
by the reconnection electric field $E_y$ \cite{zeni01}.
Figure \ref{fig:PSD}(b) presents
the electron phase-space diagram in $u_y$--$z$,
integrated over $495 \le x \le 505$.
The electrons are accelerated in $-y$ near the midplane $z \approx 450$. 
They travel in $-y$
through the Speiser motion \cite{speiser65}, bouncing in $z$.
Interestingly,
the energetic electrons tend to be concentrated around the midplane.
This is consistent with the relativistic feature of the Speiser motion.
The Speiser $z$-bounce motion is a damped oscillation in $z$.
Speiser \cite{speiser65} showed 
that the bounce width in $z$ decays like $z_{\rm max} \sim t^{-1/4}$
and
that the kinetic energy is proportional to $\varepsilon\sim t^2$
in a simple 1D configuration.
Thus we expect $z_{\rm max} \sim \varepsilon^{-1/8}$.
In contrast, in the relativistic regime,
it was recently shown that
the bounce width shrinks much faster than in the nonrelativistic regime,
$z_{\rm max} \sim \gamma^{-1/3} \sim \varepsilon^{-1/3}$ \cite{uz11c}. 
Since the high-energy electrons are directed to $-y$ (Fig.~\ref{fig:PSD}(a)),
we approximate $|u_y| \approx \varepsilon$, and then
we expect $z_{\rm max} \propto \varepsilon^{-1/3} \approx (-u_y)^{-1/3}$,
as indicated by the thin white lines in Figure \ref{fig:PSD}(b).
The phase-space distribution is in agreement with the relativistic shrinkage.
The high-energy electrons are slightly more confined in $z$ than predicted, but
this is probably because
the reconnected magnetic field $B_z$ ejects particles
in the $\pm x$ directions, and
because the system is still developing.

The inner bipolar layer of $Q_{eyz}$ (Fig.~\ref{fig:PQ}b)
corresponds to the typical bounce width in $z$ (Fig.~\ref{fig:PSD}b).
Inside this meandering channel,
there exist a huge amount of the high-energy electrons,
as evident in the VDF (Fig.~\ref{fig:PSD}a).
These energetic electrons lead to
the heat flow in the $-y$ direction.
They follow the meandering motion and then scatter
the $-y$-momentum inside the meandering channel:
upward in the upper half ($Q_{eyz}<0$) and
downward in the lower half ($Q_{eyz}>0$).
Meanwhile, in the outer $Q_{eyz}$ layer,
the electrons travel from the upstream regions toward the midplane, and then
start to be accelerated in $-y$ as they approach the midplane.
In addition, some meandering electrons travel through the outer $Q_{eyz}$ layer, but
their orbits shrink to the midplane as they are accelerated in $-y$.
As a result, the $-y$-momentum of the energetic electrons are transported
toward the meandering channel:
downward in the upper half ($Q_{eyz}>0$) and
upward in the lower half ($Q_{eyz}<0$) (Fig.~\ref{fig:PQ}(b)). 

The spacial gradient of the heat-flow tensor results in
the heat-flow inertial effect in the Ohm's law (Eq.~\ref{eq:ohm}).
Around the midplane, between the inner $Q_{eyz}$ layers,
the outward transport of the current-carrying heat-flow leads to
the inertial effects in the form of $-\partial_z Q_{eyz}>0$
(Fig.~\ref{fig:ohm}).
The underlying physics is essentially the same as
the momentum diffusion in the $-\partial_z P_{eyz}>0$ term \cite{hesse11}.
Outside there, between the inner and outer $Q_{eyz}$ layers,
the heat flows in the $-y$ direction are transported
from the midplane and from the inflow regions.
This results in the convergence of the current-carrying heat-flow, and therefore
the heat-flow inertial term has the opposite sign, $-\partial_z Q_{eyz}<0$.
Farther outside, between the outer $Q_{eyz}$ layers and the inflow regions
($430 \lesssim z \lesssim 435$ and $465 \lesssim z \lesssim 470$),
the divergence of the current-carrying heat-flow again
leads to an effective inertial effect in the form of $-\partial_z Q_{eyz}>0$.
The heat-flow inertial term changes the sign 5 times,
because of the 4-layer structure of $Q_{eyz}$.

In this case,
the heat-flow inertial term ($\nabla \cdot {Q}^{ij}$) plays a smaller role than
the divergence of the pressure tensor term ($\nabla \cdot {P}^{ij}$) to
sustain the reconnection electric field $E_y$ at the X line.
This is reasonable, because
the number of energetic electrons are much smaller than
that of the core electrons (Figs.~\ref{fig:PSD}(a,b)).
However, importantly,
the heat-flow inertial term is nonzero and
it plays a similar role as the pressure-tensor term.
It remains nonzero at the midplane
across the entire reconnection site ($370\lesssim x \lesssim 630$),
as evident in the relevant $Q_{eyz}$ layers (Fig.~\ref{fig:PSD}b).

\section{Energy balance}
\label{sec:balance}

Using the Eckart decomposition,
we further study the energy balance during the reconnection process.
The energy balance is described by the $\alpha=0$ part of the following equation,
\begin{equation}
\partial_{\beta}(T^{\alpha\beta}_{(p)}+T^{\alpha\beta}_{(e)}+T^{\alpha\beta}_{(EM)})=0
,
\end{equation}
where $T^{\alpha\beta}_{(EM)}$ is the stress-energy tensor by the electromagnetic field.
We evaluate the energy flux part (the $0i$ components) of
the plasma stress-energy tensor,
\begin{eqnarray}
\label{eq:eflux}
T^{0i}
&=& \mathcal{E} u^{0}u^{i} + P^{0i} + Q^{0i} \\
\label{eq:eflux2}
&=&
N^i + (\gamma-1) N^i + \Big[ (\mathcal{E} - n)u^{0}u^{i} + P^{0i} \Big] + Q^{0i}.
\end{eqnarray}
In Equation \ref{eq:eflux2},
we rewrite the right hand side
by using the matter flow $N^i=nu^i$.
The second term in the right hand side
corresponds to the bulk kinetic energy flux
in the nonrelativistic case.
The third term between the square brackets is
a combination of
the flow of the plasma internal energy and
the work associated with the pressure.
Thus this corresponds to the enthalpy flux.
The last heat-flow term carries an additional energy flow.
The $0i$ part of the electromagnetic stress-energy tensor is
the Poynting flux. 
In our PIC simulation,
we consider a square box at the center ($(x,z) \in [425,575] \times [410,490]$),
as indicated in Figure \ref{fig:overview}(a).
Then we integrate
the inflow energy flux ($\pm [T^{0z}_{(p)}+T^{0z}_{(e)}+T^{0z}_{(EM)}]$)
along the upper and bottom sides
and
the outflow energy flux ($\pm [T^{0x}_{(p)}+T^{0x}_{(e)}+T^{0x}_{(EM)}]$)
along the left and right sides.

\begin{figure}[htbp]
\centering
\includegraphics[width={\columnwidth},clip]{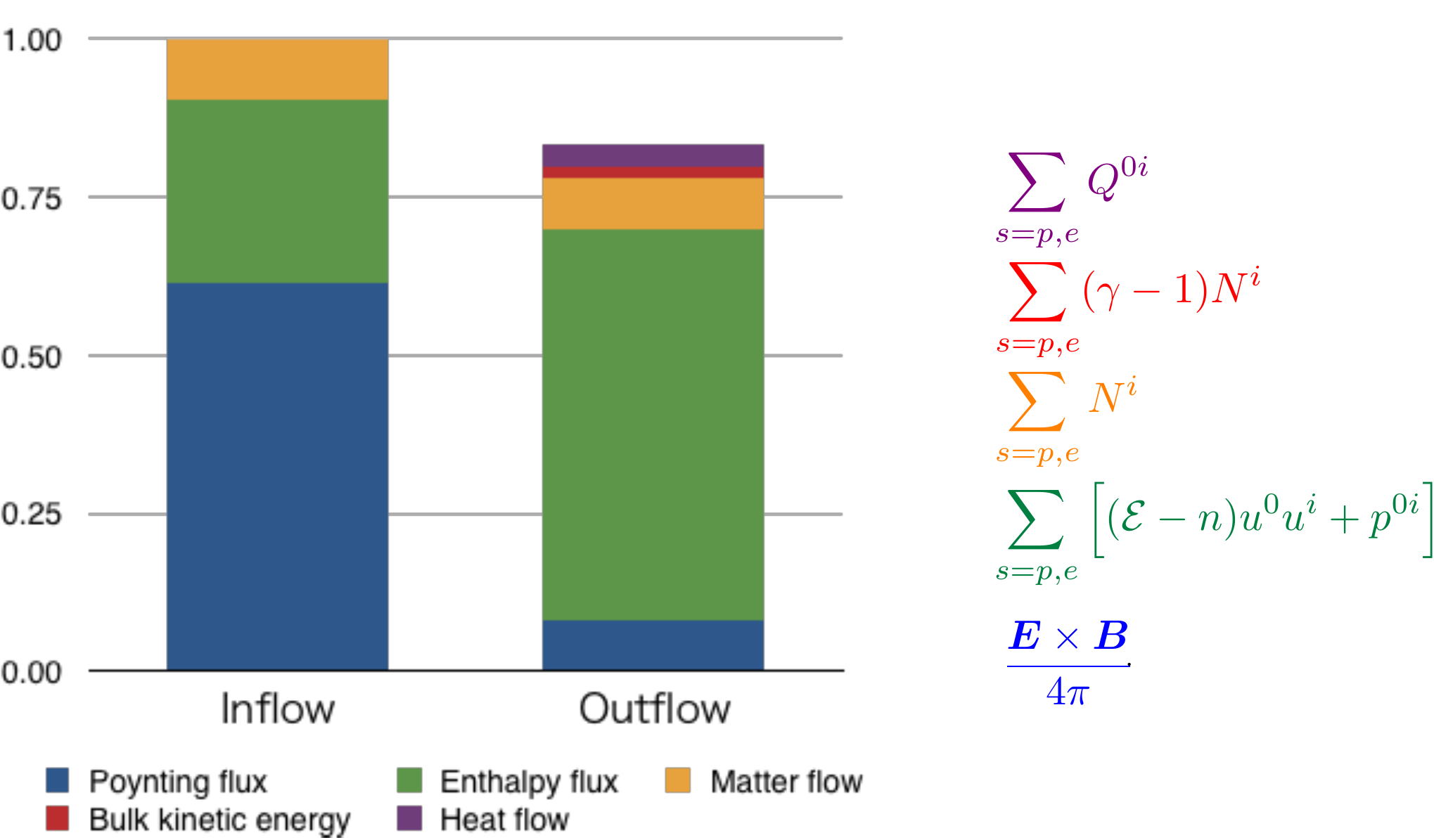}
\caption{
The composition of the incoming and outgoing energy fluxes
around the reconnection site.
\label{fig:ene}}
\end{figure}

Figure \ref{fig:ene} shows the composition of the energy fluxes at $t=85$.
The positron and electron fluxes are combined,
because they are essentially equal.
The fluxes are normalized by the total inflow energy flux,
which is larger than the total outflow flux at this point. 
The Poynting flux is the main energy carrier
in the inflow region.
The enthalpy flux is the second carrier.
This is larger than the matter flow,
because the plasma temperature is initially relativistic $T=1$.
The classical bulk kinetic energy and the heat flow are negligible.

The composition of the outflow energy flux is
very different from that of the inflow flux.
The enthalpy flux is the biggest energy carrier.
This tells us that
the reconnection process convert
the electromagnetic energy in the upstream region
into the plasma heat and
that the relativistically hot plasma carries
the energy in the form of the enthalpy flux
in the downstream region.
The matter flow is comparable with that in the inflow region.
The bulk kinetic energy flux carries some amount of the energy. 
In this case, as stated before,
plasmas inside the reconnection layer become
relativistically hot $T \gtrsim 1.5$,
and therefore
the enthalpy flux exceeds
the combination of the matter flow and the bulk kinetic energy flux,
i.e., $4 nT u^0u^i \gg nu^0u^i$. 
The heat flow carries more energy than the bulk kinetic energy flux.
The Poynting flux is largely carried by the oblique magnetic field lines
away from the midplane.

%

\section{Discussion and Summary}
\label{sec:discussion}

We have revisited
the composition of the reconnection electric field
in relativistic pair plasma reconnection.
Using the Eckart decomposition \cite{ec40},
we have formulated the relativistic kinetic Ohm's law.
Analyzing the PIC simulation data,
we have basically confirmed the conventional picture that
the divergence of the pressure tensor balances
the reconnection electric field,
in agreement with nonrelativistic studies \cite{cai97,hesse99,hesse11}
and
the previous study with different formulation \cite{hesse07}.
We have further found
that the relativistic temperature enhances the bulk inertial effect
and
that the new heat-flow inertial effect arises from
the Speiser motion of the energetic electrons. 
Note that the heat flow is different from the heat conduction,
because the inertial effect arises from the off-diagonal components of $Q$,
while the heat conduction is usually driven by the temperature gradient,
i.e., $\nabla T$. The heat conduction is unable to sustain
the reconnection electric field $E_y$ in the 2D system of $\partial_y=0$.
Additional analysis on the VDF, the phase-space diagram, and
the spatial distribution of $Q_{eyz}$ suggests that
the heat-flow inertial effect stems from
the Speiser particle acceleration around midplane,
which is a robust feature of relativistic reconnection.
In this sense, the underlying physics of the new term is similar to
that of the conventional pressure-tensor term,
which stems from the Speiser motion of the bulk populations. 
The 4-layer structure of $Q_{eyz}$ leads to
the characteristic profile of the heat-flow inertial term.


We have also studied the energy balance during the reconnection process.
We have found that
the incoming Poynting flux is transferred to
the enthalpy flux in the outflow region.
The heat flow carries more energy than the matter flux in the outflow region,
however, the enthalpy flux is a major carrier of the outgoing energy.
This is consistent with the previous two-fluid simulation \cite{zeni09b}
and extends previous nonrelativistic results \cite{aunai11c,yamada15,yamada16,lapenta17}.
It is reasonable that
the upstream magnetic energy is transferred to
the plasma internal energy, a energy of a high-entropy state,
through a dissipative process like magnetic reconnection.

There are some limitations in our analysis.
First, our system is still evolving in time.
The temporal part is non-negligible in the Ohm's law (Fig.~\ref{fig:ohm}),
and
the inflow and outflow energy fluxes are not equal in the energy balance (Fig.~\ref{fig:ene}).
Second, our box size may not be ideal to discuss the energy balance,
because it is shorter in $x$ than the extent of the dissipation region
(Figs.~\ref{fig:overview}a,b).
These two issues may be overcome by using a larger simulation domain.
In a larger domain, the system might reach a quasisteady stage,
so that we can rule out time-dependent effects.
The outflow speed might reach the upstream Alfv\'{e}n speed,
as suggested by the theory \cite{lyub05} and MHD simulations \cite{zeni10b}.
Then, the bulk kinetic energy flux will carry
a larger amount of the energy in the outflow region. 
However, even in this case,
the reconnection site started to generate secondary islands after $t \gtrsim 85$.
In our experience,
the number of particles in a cell often controls the onset of the island formation,
but we used a large number of particles in this study. 
Considering that larger PIC simulations often exhibit secondary islands,
it may not be necessary to stick to the quasi-steady picture. 
%
%
Third, we have analyzed only one run.
In astrophysical applications,
we expect that the reconnection occurs in extreme situations,
in which the initial magnetic energy vastly exceeds the plasma energy. 
The upstream leptons are often supposed to be cold,
due to the synchrotron cooling. 
In such extreme cases,
since more energy will be transferred to energetic particles,
and since their energy easily exceeds the typical thermal energy,
the heat-flow terms could be more pronounced,
both in the Ohm's law and in the energy balance.
For example, if the heat-flow inertial term replaces the pressure-tensor term in the Ohm's law,
a widespread belief that
the pressure-tensor term is essential at the X line \cite{vas75,hesse11}
will be challenged.
These issues are left for future investigations.


We have employed
the Eckart velocity as the fluid velocity in this work.
If we employed Landau velocity ($u^\alpha_{(L)}$) instead,
since $q^\alpha=0$ in Equation \ref{eq:eckart},
the heat-flow inertial term in the Ohm's law and 
the heat-flow term in the energy flux disappear.
However, the Eckart velocity ($u^\alpha_{(E)}$) should be used
in the right hand side of the energy-momentum equation (Eq.~\ref{eq:T}),
because the plasma number flow carries the electric current.
To evaluate the Ohm's law (Fig.~\ref{fig:ohm}),
we have to use both $u^\alpha_{(E)}$ and $u^\alpha_{(L)}$, or
to introduce an additional flow,
$u^\alpha_{(C)}\equiv u^\alpha_{(E)}$-$u^\alpha_{(L)}$.
In the energy balance,
it is less meaningful to discuss the matter flow $N^i=nu^i_{(E)}$ and
relevant quantities by using the Landau-frame quantities.
The energy dissipation (Eq.~\ref{eq:DeN}) also requires
{\it the number density's flow} \cite{zeni11c}, i.e., the Eckart velocity. 
Based on these considerations,
we prefer the Eckart-frame approach to evaluate the PIC results.

In summary, using the Eckart decomposition,
we have analyzed the kinetic Ohm's law and the energy balance
in relativistic reconnection in a pair plasma.
We have basically confirmed previous expectations that
the pressure-tensor term is responsible for
the reconnection electric field $E_y$ \cite{hesse07}
and that the relativistic enthalpy-flux carries the outgoing energy,
but additional roles by the relativistic heat-flow terms are discovered.
In addition to these results,
it is meaningful to demonstrate
the diagnostics for the relativistic PIC data.
Then it will be possible to cross-compare
relativistic PIC simulations, relativistic fluid simulations, and relativistic MHD theories.


~\\
{\bf Acknowledgments}\\
This work was supported by
Grant-in-Aid for Young Scientists (B) 25871054 and
Grant-in-Aid for Scientific Research (C) 17K05673
from the Japan Society for the Promotion of Science (JSPS).

\appendix

\section{Eigenvalue properties for the Landau frame}
\label{sec:Landau}

Since the energy flow disappears in the Landau frame,
$q^{\mu}(u^\mu_{(L)})=0$,
the Landau-frame four-velocity $u^\mu_{(L)}$ satisfies
\begin{equation}
\label{eq:eigen}
T^{\alpha}_{\delta} u^{\delta}_{(L)}
=
(-u_{(L)\beta} T^{\beta\gamma} u_{(L)\gamma})u^{\alpha}_{(L)}
.
\end{equation}
This indicates that $u^\mu_{(L)}$ is
an eigenvector of a 4$\times$4 matrix $(T^{\alpha}_{\delta})$.
The factor on the right hand side is equivalent to
the energy density in the Landau frame,
and therefore the eigenvalue is negative,
$-\mathcal{E}(u^{\alpha}_{(L)})<0$.


Let us further consider a space-like vector
$x'^{\alpha} = (0, x'^i)$ in the Landau frame.
Since $T'^{i0}=T'^{0i}=0$, there exists three vectors
that satisfy the following relation for a scalar $p'$,
\begin{equation}
\label{eq:p}
T'^{\alpha\beta}x'_{\beta} = p'^{\alpha\beta}x'_{\beta} = p' x'^{\alpha}
,
\end{equation}
where $p'=p'_{1,2,3}$ and $x'=x'_{1,2,3}$ are
the eigenvalues and eigenvectors of the pressure tensor.
Transforming Equation \ref{eq:p} to the observer frame,
\begin{equation}
\Lambda^{\alpha}_{\mu} T'^{\mu\beta}g_{\beta\nu}x'^{\nu}
= p' (\Lambda^{\alpha}_{\mu} x'^{\mu})
= p' x^{\alpha}
\end{equation}
Here we used $g_{\beta\nu}x'^{\nu}=x'_{\beta}$,
because $x'$ has no time component.
We also use the definition of the Lorentz transform,
${\Lambda}_{\beta}^{\gamma}g_{\gamma\delta}{\Lambda}_{\nu}^{\delta}=g_{\beta\nu}$,
\begin{equation}
\Lambda^{\alpha}_{\mu} T'^{\mu\beta}
{\Lambda}_{\beta}^{\gamma}
g_{\gamma\delta}
{\Lambda}_{\nu}^{\delta}
x'^{\nu}
=
T^{\alpha}_{\gamma} x^{\gamma}
=
p' x^{\alpha}
.
\end{equation}
This tells us that 
Equation \ref{eq:eigen} has
three positive eigenvalues and the space-like eigenvectors,
corresponding to the eigenvalues of the pressure tensor and the spatial axises
in the Landau frame.  From their definitions in the Landau frame,
\begin{equation}
\mathcal{E}'\equiv
\int f'(u')\gamma' \gamma' \frac{d^3u'}{\gamma'} 
>
\int f'(u')u'^2 \frac{d^3u'}{\gamma'} 
=
p'_1 + p'_2 + p'_3
,
\end{equation}
it is obvious that
the negative eigenvalue $-\mathcal{E}'$ has the biggest absolute value among the four eigenvalues.

In addition, the energy density in the observer frame yields
\begin{equation}
T^{00} = \Lambda^0_\alpha\Lambda^0_\beta T'^{\alpha\beta}
= \gamma^2 \mathcal{E}' + (P'\vec{u}_{(L)})\vec{u}_{(L)}
\end{equation}
where $\vec{u}_{(L)}=(u^1,u^2,u^3)$ is the spatial part of the Landau four-velocity.
By rewriting $\vec{u}_{(L)}=a_1\vec{x'}_1+a_2\vec{x'}_2+a_3\vec{x'}_3$ with the unit eigenvectors $\vec{x'}_{1,2,3}$,
we find
\begin{eqnarray}
T^{00} = (1+a_1^2+a_2^2+a_3^2) \mathcal{E}' + (p'_1 a_1^2 + p'_2 a_2^2 + p'_3 a_3^2) \ge \mathcal{E}'
.
\end{eqnarray}
This tells us that the energy density always increases,
by the Lorentz transformation from the Landau frame.
In other words, the Landau frame is a frame that minimizes the energy density.

~\\
{\bf References}\\

\end{document}